\documentstyle[12pt]{article}

\title{Separation of \mbox{$S$--wave} pseudoscalar and
pseudovector amplitudes in \mbox{$\pi^- p_{\uparrow} \rightarrow
\pi^+ \pi^- n$} reaction on polarized target}
\author{R. Kami\'nski, L. Le\'sniak and K. Rybicki \\
Henryk Niewodnicza\'nski Institute of Nuclear Physics, \\ PL 31-342 Krak\'ow, 
Poland}

\newcommand{\be}{\begin{equation}}
\newcommand{\ee}{\end{equation}}
\newcommand{\ba}{\begin{eqnarray}}
\newcommand{\ea}{\end{eqnarray}}
\newcommand{\pipi}{$\pi\pi$ }
\newcommand{\KK}{$K\overline{K}$ }
\newcommand{\fo}{$f_0(980)$ }
\newcommand{\ro}{$\rho(770)$ }
\newcommand{\fd}{$f_2(1270)$ }
\newcommand{\rf}{$\rho_3(1690)$ }

\newcommand{\aone}{$a_1$ }
\newcommand{\ep}{$f_0(750)$ }
\newcommand{\fgg}{$f_0(1500)$ }
\newcommand{\mpp}{$m_{\pi\pi}$ }
\newcommand{\g}[1]{$\overline{g}_{#1}$ }
\newcommand{\h}[1]{$\overline{h}_{#1}$ }
\newcommand{\del}{$\delta$ }
\newcommand{\et}{$\eta$ }
\newcommand{\sig}{$\sigma$ }
\newcommand{\downp}{"down--flat" }
\newcommand{\downs}{"down--steep" }
\newcommand{\upp}{"up--flat" }
\newcommand{\ups}{"up--steep" }
\newcommand{\crc}{$^{\circ}$ }

\begin{document}
\maketitle
\begin{abstract}
 
A new analysis of $S$--wave production amplitudes for the reaction
$\pi^- p_{\uparrow} \rightarrow \pi^+ \pi^- n$ on a transversely polarized target
is performed.
It is based on the results obtained by the CERN--Cracow--Munich collaboration
in the 
\pipi energy range from 600 MeV to 1600 MeV at 17.2 GeV/c $\pi^-$ momentum.
Energy--independent separation of the $S$--wave pseudoscalar amplitude 
($\pi$ exchange) from the pseudovector amplitude (\aone exchange) is
carried out using assumptions much weaker than those in all previous analyses. 
We show that, especially around 1000 MeV and around 1500 MeV,
the \aone exchange amplitude cannot  be neglected. 
The scalar--isoscalar $\pi\pi$ phase shifts 
are calculated using fairly  weak assumptions. 

Below the $K\overline{K}$ threshold we find two solutions for the 
 $\pi-\pi$ phase shifts, for which the phases increase slower with the 
 effective $\pi-\pi$ mass than the P-wave phases. Both solutions are consistent 
 with a broad
$f_{0}(500)$ but only one is similar to the well-known "down" solution.
We find also the third solution (with a somewhat 
puzzling behavior of inelasticity) which exhibits  a narrow $f_{0}(750)$ 
claimed by Svec. All the solutions undergo a rapid change at the 
$K\overline{K}$ threshold. Above 1420 MeV the phase shifts increase with
energy faster than those obtained without the polarized-target data. This
phase behavior as well as an increase  of the modulus of the $a_{1}$-exchange
amplitude can be due to the presence of the  $f_{0}(1500)$.

\end{abstract}
 
 
\section{Introduction \label{introd}}
 
\hspace{0.6cm}
Study of scalar mesons is one of the central points of light 
quark spectroscopy. 
In addition to ordinary $q\overline{q}$ mesons, some \KK bound states 
\cite{w-i83,klm} 
and lowest--lying glueballs are expected as well 
\cite{anisovich,amsler_glasgow,amsler_close96,ukqcd,th_groups}.
Unfortunately, the experimental situation is still far from being clear 
\cite{trnqv_conf,pen_hadron95,bugg_hadron95,resag_hadron95}. 
In the effective mass region above 1000 MeV,
a rich spectrum of scalar mesons has been recently proposed and discussed in 
many experimental
\cite{exp_groups1,exp_groups3} 
and theoretical 
\cite{ukqcd,th_groups,tornqvist_zp95} 
papers.
Generally, the proposed scalar states have a coupling to 
the \pipi channel strong enough to manifest themselves through energy 
dependence of the \pipi interaction amplitudes.
In the past few years  scalar resonance \fgg related to a hypothetical lowest 
lying glueball state was announced 
\cite{anisovich,amsler_glasgow,amsler_close96,anisovich_plb364}.
Thus, a study of the \pipi interaction near 1500 MeV is important also 
in this context. Below 1000 MeV the status of scalar mesons  is also unclear 
since the existence of a broad $\sigma$  or $f_0(750)$ 
meson  still remains an open question
\cite{klm,svec,svec95,torn96}.
 
One of the main sources of information on the production of scalar states
is the \pipi partial wave analysis (PWA) yielding the $S$--wave.
It should be stressed that study of $S$--wave objects {\it does require} the 
partial wave analysis to "subtract" contributions of leading mesons \ro, \fd  
and \rf  which dominate the total cross section. 
Virtually all PWA's were based on the old  CERN--Munich experiment
 \cite{grayer} 
which supplied 3$\times$10$^5$ events of the reaction
 
\be
\pi^-p\rightarrow \pi^+\pi^- n
\label{reaction}
\ee
at 17.2 GeV/c. The number of observables provided by
such experiment is much smaller than the number of real parameters needed to 
describe the partial waves. Consequently, the dominance of pseudoscalar 
exchange, equivalent to the absence of pseudovector exchange
 and several other physical assumptions have been made
in previous studies
\cite{grayer}--\cite{ochsnc}.
 
In this paper we use results of PWA performed in the energy range from 
600 MeV to 1600 MeV (in 20 MeV  bins) obtained with the help of 
the  polarized target experiment. This experiment, performed 20 years ago
by  the CERN--Cracow--Munich collaboration,  provided 1.2$\times$10$^6$ events 
of the reaction
 
\begin{equation}
\pi^-p_{\uparrow}\rightarrow \pi^+\pi^- n
\label{reaction_pol}
\end{equation}
also at 17.2 GeV/c \cite{becker}. 
Combination of results of both experiments  yields a number of 
observables sufficient for performing a 
quasi--complete and energy independent PWA without any model assumptions. 
This analysis is only quasi--complete  because of an unknown phase 
between  two sets of transversity amplitudes. 
Nevertheless, intensities of partial waves could
be determined in a completely model--independent way. 
This removed  
ambiguities appearing in  earlier studies, except for the old "up--down" 
ambiguity 
\cite{mms}. 
The "up" solution contains an $S$--wave resonance  just under
the \ro and of similar width, while the "down" $S$--wave modulus stays high and 
nearly constant all the way to the \fo. It was only Svec\cite{svec,svec95} 
who argued  persistently in favour of the "up" solution,  using both the 
17.2 GeV data as well as data on the reaction 
$\pi^+p_{\uparrow}\rightarrow \pi^+\pi^- p$ at 5.85 and 11.98 GeV/c.
However,  general belief (see e.g. 
\cite{pen&protop},
\cite{morgan}) 
was  that the "up--down" ambiguity had been resolved definitely in favour of 
the "down" solution. We disagree with this belief since  all the previous 
studies of the {\it full} 17.2 GeV/c data were consistent with both the "up" 
and "down" solutions. The same is true for the present analysis. We stress 
this point because the mini--reviews in the last RPP editions (see e.g. Ref.
\cite{pdg96}) 
contained the sentence "BECKER 79B \cite{becker_b} excluded a narrow 
resonance behaviour for $\delta^0_0$ ..." , contrary to what is 
stated explicitly in another paper of the same collaboration  
(see  Sect. 6 of  \cite{becker}): "...our results do not give a clear 
answer to this ambiguity...".

In this paper we make another step in the analysis of 17.2 GeV/c data 
attempting to bridge two sets of transversity amplitudes. The phase of each 
$S$--wave transversity amplitude is fixed by requiring the phase of the 
leading $P$, $D$, $F$--waves to follow roughly the phase of the Breit--Wigner 
amplitudes of the \ro, \fd and \rf resonances in the low, medium and high 
mass region respectively. 
This fairly reasonable assumption  allows us to 
separate explicitly for the first time the pseudoscalar and pseudovector 
amplitudes in the $S$--wave.
 
In Sect. 2 we present mathematical formalism needed to separate the 
one--pion and \aone exchange amplitudes for the reaction on a polarized target.
Section 3 contains a short description of the PWA done by the CERN--Cracow--Munich 
collaboration for  reaction (\ref{reaction_pol}). In Sect. 4 we present 
our analysis yielding pseudoscalar and pseudovector reaction amplitudes.
Further on, from the pseudoscalar amplitude we extract the $I=0$, $S$--wave 
amplitude describing the \pipi elastic scattering amplitude. Our results are 
discussed in Sect. 4 and  summarized in Sect. 5.

 
\section{
Amplitudes describing pion pair production in the \mbox{\boldmath $I=0$}, 
\mbox{\boldmath $S$}--wave 
\label{form}}
 
{\bf a) Separation of pseudoscalar and pseudovector exchange amplitudes}
 
\vspace{0.3cm}
 
\hspace*{6mm}Let us denote by $f_0$ a system of two pions in a relative
$S$--wave isospin 0 state. Transition amplitude for the $f_0$ production
 process
$\pi^{-}p\rightarrow f_0~n$ can be written as the following matrix
element
\be
T_{s_p s_n}=<u^{s_n}_{p_2}|A\gamma_{5}+ \frac{1}{2}B\gamma_{5}
\gamma_{\mu}(p_{\pi}+p_f)^{\mu}|u^{s_p}_{p_1}>,         \label{1}
\ee
where $p_1,p_2,p_{\pi}$ and $p_f$ are proton, neutron, incoming pion, and
final $f_0$ four-momenta, $s_p$ and $s_n$ are proton and neutron spin
projections, $u^{s_p}_{p_1}$ and $u^{s_n}_{p_2}$ are the corresponding 
four-spinors,
A and B are functions of the Mandelstam variables $s=(p_1+p_{\pi})^2$
and $t=(p_1-p_2)^2$ at fixed $f_0$ mass $m_{\pi\pi}$. Part A of the amplitude 
corresponds to the pseudoscalar (or one pion) exchange while part B describes
the pseudovector exchange which we shall briefly call $a_1$ exchange
since we expect that the $a_1$ meson exchange amplitude constitutes its
major contribution. Functions A and B have to be determined from 
experiment.
Using s-channel helicity amplitudes in the c.m. $\pi^{-} p$ system one can
derive the following two independent amplitudes   
\cite{mms}:
\ba
H_{++}\equiv T_{{1\over 2}{1\over 2}}=
            - T_{-{1\over 2}-{1\over 2}},    \label{2} \\
H_{+-}\equiv T_{{1\over 2}-{1\over 2}}=
             T_{-{1\over 2}{1\over 2}},    \label{3}  \\
H_{++}=(\frac{A}{2M}\sqrt{-t_{min}}-\frac{B}{2} 
\frac{m^2_f-m^2_{\pi}}{\sqrt{-t_{min}}})\cos\frac{\Theta_s}{2},
                                                             \label{4} \\
H_{+-}=(-\frac{A}{2M}\sqrt{-t_{max}}+\frac{B}{2} 
\frac{m^2_f-m^2_{\pi}}{\sqrt{-t_{max}}})\sin\frac{\Theta_s}{2}.
                                                             \label{5}  
\ea
In (5-6) $\Theta_s$ is the neutron scattering angle (with respect to proton
direction), $M$ and $m_{\pi}$ are nucleon (proton and neutron average) and
pion masses, $t_{min}$ and $t_{max}$ are expressed by momenta $p_1$,
$p_2$ and the corresponding c.m. energies $E_1,E_2$:
\ba
t_{min}=2(M^2+p_1p_2-E_1E_2),                         \label{6} \\
t_{max}=2(M^2-p_1p_2-E_1E_2).                         \label{7} 
\ea
The scattering angle is related to the four-momentum transfer squared $t$:
 
\be
\sin{\frac{\Theta_s}{2}}=\sqrt{\frac{t_{min}-t}{t_{min}-t_{max}}}.
                                                           \label{8}
\ee
 
In this paper we use two amplitudes $g$ and $h$, closely related to $H_{++}$ and 
$H_{+-}$, adequate for describing $f_0$
production on a transversely polarized target: 

\ba
g\equiv<n\downarrow|T|p\uparrow>=(H_{+-}-iH_{++})\,exp(\frac{1}{2}i
\Theta_s),                                                 \label{9}   \\
h\equiv<n\uparrow|T|p\downarrow>=(H_{+-}+iH_{++})\,exp(-\frac{1}{2}i
\Theta_s).                                                 \label{10}   
\ea
In this case, nucleon spin is quantized along the vector 
$\bf{n}=\bf{p_1\,\times\,p_2}$ normal to the production plane (in (11) and (12),
arrows
$\uparrow$ and $\downarrow$ denote spin projections parallel or
 antiparallel 
to {\bf n}). It can be shown that the remaining two matrix elements vanish:
\be
<n\uparrow|T|p\uparrow>=<n\downarrow|T|p\downarrow>=0.     \label{11}
\ee
 
Separation of invariant amplitudes $A$ and $B$ for the reaction
$\pi^{-}p\rightarrow f_0~n$ for {\it fixed} values of t and $m_{\pi\pi}$ can be done
in the following way. Using (6,7) and (11,12) we define components 
$g_A$, $g_B$ and $h_A,h_B$ of amplitudes g and h as follows:
\ba
g\equiv g_A+g_B, \:\:\:\: g_A\equiv \frac{A}{2M}~U\:\:\:\: 
g_B\equiv \frac{B}{2}rV,                                   \label{12} \\
h\equiv h_A+h_B, \:\:\:\: h_A\equiv \frac{A}{2M}~U^*\:\:\:\:
h_B\equiv \frac{B}{2}rV^*,                                   \label{13}
\ea
where
\ba
U=(b-a)cs+i(bc^2+as^2),                                 \label{14} \\
V=(1/a-1/b)cs -i(s^2/a+c^2/b)                           \label{15} 
\ea
with the following shorthand notation: $a=\sqrt{-t_{max}},~b=\sqrt{-t_{min}},\\
r=m^2_f-m^2_{\pi},~c=\cos(\frac{1}{2}\Theta_s),
~s=\sin(\frac{1}{2}\Theta_s)$.
 
From (14,15) one can obtain the desired amplitudes $\frac{A}{2M}$
and $\frac{B}{2}r$:
\ba
\frac{A}{2M}=\frac{1}{W}(gV^*-hV),                           \label{16}\\
\frac{B}{2}r=\frac{1}{W}(-gU^*+hU),                          \label{17}
\ea
where
\be
W=UV^*-U^*V.                                                  \label{18}   
\ee
 
Solutions (18,19) are well suited for the analysis of experimental data
taken with narrow t-bins. The CERN-Cracow-Munich data \cite{becker} have
been, however, averaged over a relatively wide $t$-range:
\be
t_1=-0.2 \mbox{ (GeV/c)$^2$}<t<t_2=-0.005 \mbox{ (GeV/c)$^2$}.                    \label{19}
\ee
Therefore, we need amplitudes $\overline{g}$ and $\overline{h}$
averaged over this $t$-range,
where
\be
\overline{g} \equiv \frac{1}{t_2-t_1} \int_{t_1}^{t_2}g(t)~dt   \label{20}
\ee
with similar equation for $\overline{h}$. Then according to (14-15)
we write
\ba
\overline{g}=\frac{1}{2M}\overline{AU}+\frac{1}{2}r\overline{BV} \label{21}\\
\overline{h}=\frac{1}{2M}\overline{AU^*}+\frac{1}{2}r\overline{BV^*},
                                                                   \label{22}
\ea
where symbols $\overline{AU},\overline{BV}$ etc. are defined as in (22).
 
In order to proceed further we have to assume some $t$-dependence of 
functions $A$ and $B$ allowing for complete freedom in the effective mass 
dependence. Thus, we write
\be
A(m_{\pi\pi},t)=A_0(m_{\pi\pi})p(m_{\pi\pi},t)                                          \label{23}
\ee
and
\be
B(m_{\pi\pi},t)=B_0(m_{\pi\pi})q(m_{\pi\pi},t),                                          \label{24}
\ee
where $p(m_{\pi\pi},t)$ and $q(m_{\pi\pi},t)$ are postulated functions of $t$ (which may
also depend on $m_{\pi\pi})$, $A_0(m_{\pi\pi})$ and $B_0(m_{\pi\pi})$ depend on $m_{\pi\pi}$ only  and
 their
behaviour should be determined from experiment. A possible parametrization
of $p(m_{\pi\pi},t)$ is:
\be
p(m_{\pi\pi},t)=\frac{e^{a(m_{\pi\pi})t}}{m_{\pi}^2-t},                       \label{25}
\ee
which is equal to the pion propagator multiplied by the exponential form
factor with parameter $a(m_{\pi\pi})$ being an  a priori unknown function of $m_{\pi\pi}$.
This shape of functional dependence
 has been introduced in many pion exchange models.
For $q(m_{\pi\pi},t)$ we have used four different parametrizations:~ 1,~$t$,~$exp(bt)$ and
~$t\,exp(bt)$ with $b$=4 GeV$^{-2}$. 
As we shall show later, separation of averaged amplitudes $\overline{g}$ and $\overline{h}$ into a sum of  
averaged parts $\overline{g_A}+\overline{g_B}$ and $\overline{h_A}+
\overline{h_B}$ as in (14-15) and (23-24) is largely insensitive to the
form of parametrization of function $q(m_{\pi\pi},t)$. 
Functions
$\overline{g_A},\overline{h_A}$, $\overline{g_B}$ and $\overline{h_B}$
are linear combinations of amplitudes $\overline{g}$ and $\overline{h}$:
\ba
\overline{g_A}=c_1\overline{g}+c_2\overline{h},                \label{26}\\
\overline{h_A}=d_1\overline{g}+d_2\overline{h},                \label{27}\\
\overline{g_B}=d_2\overline{g}-c_2\overline{h},                \label{28}\\
\overline{h_B}=-d_1\overline{g}+c_1\overline{h},                \label{29}
\ea
where the complex coefficients are
\ba
c_1=\overline{qV^*}\:\overline{pU}/D,                            \label{30} \\
c_2=-\overline{qV}\:\overline{pU}/D,                             \label{31}  \\ 
d_1=\overline{qV^*}\:\overline{pU^*}/D,                             \label{32}\\
d_2=-\overline{qV}\:\overline{pU^*}/D                             \label{33}  
\ea
and
\be
D=\overline{pU}\:\overline{qV^*}-\overline{pU^*}\:\overline{qV}.  \label{34}
\ee
Having functions $\overline{g}$ and $\overline{h}$ experimentally
determined, Eqs (28-36) make it possible to calculate the unknown functions $A_0(m_{\pi\pi})$
and $B_0(m_{\pi\pi})$ in the following way:
\be
\frac{A_0(m_{\pi\pi})}{2M}=\frac{\overline{g_A}}{\,\overline{pU}}           \label{35}
\ee
and
\be
\frac{B_0(m_{\pi\pi})}{2}r=\frac{\overline{g_B}}{\,\overline{qV}}.           \label{36}
\ee
From (25-26) we then obtain the desired functions $A$ and $B$, so that separation
of amplitude (3) into pseudoscalar and pseudovector parts is finally
achieved.
 
\vspace{0.3cm}
 
{\bf b) Determination of scalar-isoscalar pion-pion interaction amplitude}
 
\vspace{0.3cm}
 
\hspace*{6mm}Separation of the pseudoscalar amplitude, dominantly corresponding
to the one pion exchange contribution to the $\pi^{-}p\rightarrow f_0~n$
process, is the first step in the determination of the scalar-isoscalar
pion-pion amplitude $a_0$. This amplitude can be calculated if both the 
$S$--wave
$\pi^+\pi^-\rightarrow \pi^+\pi^-$ amplitude $a_S$ and the $I$=2 $S$--wave
amplitude $a_2$ are known:
\be
a_0=3~a_S-\frac{1}{2}a_2                                       \label{37}
\ee
(see  \cite{mms} p.12 and notice that the $I$=1, $S$=0 contribution vanishes).
Amplitude $a_S$ is closely related to $A_0/2M$ given by (37)
since $A_0$ is a factor responsible for the $\pi\pi$ interaction in (25):
\be
a_S=- \frac{p_{\pi}\sqrt{sq_{\pi}}\,f}{m_{\pi\pi}\sqrt{2\cdot\:\frac{g^2}{4\pi}}}
\frac{A_0}{2M},                                               \label{38}
\ee
where $p_{\pi}$ is the incoming $\pi^-$ momentum in the $\pi^-p$ c.m.
frame, $q_{\pi}$ is the final pion momentum in the $f_0$ decay frame,
$g^2/4\pi=14.6$ is the pion-nucleon coupling constant, and $f$ is the
correction factor. In this factor the averaged $t$-dependence of the pion-
nucleon vertex function and the off-shellness of the exchanged pion
are included, which allows us to apply this formula to the analysis of the data
taken in a wide t-region. 

When writing (\ref{38}) we have assumed that absorption
effects due to the final state interaction of the \pipi with the outgoing neutron can
be neglected. This assumption is supported by the results obtained in 
studies of absorption effects using the Regge phenomenology (\cite{irving}).
The point is that in the case of dominance of the nucleon helicity flip 
amplitude, final state interaction effects are relatively weak. Thus, we can
expect that errors caused by those effects are smaller than experimental
errors.
 
Amplitude $a_2$ can be measured in the process 
$\pi^+~p\rightarrow \pi^+\pi^+~n$, provided the partial wave analysis is done
and a similar separation of pseudoscalar and pseudovector contributions
is performed. Although studies  of $\pi^+\pi^+$ and  $\pi^-\pi^-$ 
systems  were in the past
\cite{hoogland}, the above--mentioned separation, which requires 
polarization measurements, has never been performed. 
Therefore, we have to rely
on  determination of the $I$=2 amplitude based on the assumption that
one-pion exchange dominates in the process under discussion. 
The $I$=2 
$S$--wave amplitude has been calculated using the data of \cite{hoogland} and 
the pion-pion separable potential model previously applied to the 
description of the coupled channel $\pi\pi$ and $K\overline{K}$ $I$=0, $S$--wave 
interactions \cite{klm}. Here for the $I$=2 channel we use a very simple 
two-parameter pion-pion potential of rank one:
\be
V_{\pi}(p,p')=\lambda G(p)G(p'),                               \label{39}
\ee
where $\lambda$ is a constant and
\be
G(p)=\sqrt{\frac{4\pi}{m_{\pi}}} \frac{1}{p^2+\beta^2}         \label{40}
\ee
is a form factor with one range parameter $\beta$; $p$ and $p'$ are pion
c.m. momenta in the initial and final states respectively. The calculated phase shifts
$\delta_2$ corresponding to the amplitude 
\be
a_2=\sin{\delta_2} exp(i\delta_2)                               \label{41}
\ee
for $\Lambda \equiv \frac{\lambda}{2\beta^3}= -0.1309$ and $\beta= 3.384$~GeV
are shown in Fig. 1. The analytical form of the amplitude $a_2$ can be found
in Appendix A of \cite{klm}.
This amplitude, along with $a_S$ obtained from
(40), was used to calculate $a_0$ given by (39). 
Amplitude $a_2$, being
generally smaller than $a_S$, cannot be, however, neglected in (39) as shown
in Sect. 4 d. Amplitude $a_0$ is normalized to Argand's form:
\be
a_0=\frac{\eta e^{2i\delta}-1}{2~i},                   \label{42}
\ee
where $\delta$ is the $I$=0, $S$=0 phase shift and $\eta$ is the inelasticity 
coefficient.
   
 
\section{
Model-independent determination of partial-wave amplitudes
\label{expryb}}
 
\hspace*{6mm}This section is a short recapitulation of what was extensively
described in the old papers of the CERN-Cracow-Munich collaboration
\cite{becker,becker_b,LR,chabaud}, 
to which the reader is referred for more details.

In the analysis, $3\times 10^{5}$ events on a hydrogen 
target\cite{grayer} were combined with $1.2\times 10^{6}$ events on a polarized
(butanol) target, both for 17.2 GeV/c $\pi^{-}$. 
The former events yield  $t_{M}^{L}$
moments while the latter provide the $p_{M}^{L}$ and $r_{M}^{L}$ moments
from interactions with protons of hydrogen bound in butanol
molecule (protons in carbon and oxygen nuclei cannot be polarized).
In this way we obtain a quasi-complete description of the reaction:\\
\begin{equation}
\pi^{-}p_{\uparrow} \rightarrow \pi^{+}\pi^{-}n.
\end{equation}
In the case of a transversely polarized target the differential cross section
can be written as
\begin{equation}
\frac{d^{2}\sigma}{dm_{\pi\pi}dtd\Omega}=\sum_{L,M} t_{M}^{L} Re Y^{L}_{M}(\Omega)+   
Pcos\psi\sum_{L,M} p_{M}^{L} Re Y^{L}_{M}(\Omega)+
Psin\psi\sum_{L,M} r_{M}^{L} Im Y^{L}_{M}(\Omega),
\end{equation}
where:\\
$m_{\pi\pi}$ - effective mass of the $\pi^{+}\pi^{-}$ system,\\
$t$ - four-momentum transfer squared from the $\pi^{-}$  beam to the $\pi^{+}\pi^{-}$ 
system,\\
$ Y^{L}_{M}$ - spherical harmonics,\\
$\Omega$ - decay angles of the $\pi^{-}$ in the t-channel  $\pi^{+}\pi^{-}$
rest frame,\\
$\psi$ - polarization angle,\\
$P$ - degree of polarization.

 \vspace{-.5cm} 
 
\begin{table}[h]
\begin{center}
Table 1
\end{center}

{\small
Number of parameters and observables in partial-wave analysis,
$l$ denotes the $\pi^-$ 
orbital momentum in the $\pi^+\pi^-$ rest frame
and $m$ -- its projection.}
\begin{center}
\vspace*{0.7ex}
\begin{tabular}{||c||c|c|c||} \hline \hline
 $l_{max}$(wave)     & $1 (P)$            & $2 (D)$          & $3 (F)$   \\
 \hline \hline
Number of amplitudes ($m\leq 1$)          &  8   &     14    &     20    \\
Number of real parameters ($m\leq 1$)     & 14   &     26    &     38    \\
Number of  $t_{M}^{L}$ moments ($M\leq 2$)&  6   &     12    &     18    \\
Number of  $p_{M}^{L}$ moments ($M\leq 2$)&  6   &     12    &     18    \\
Number of  $r_{M}^{L}$ moments ($M\leq 2$)&  3   &      7    &     11    \\
Total number of moments ($M\leq 2$)       & 15   &     31    &     47    \\                  
\hline \hline
\end{tabular}
\end{center}
\end{table}
 
\hspace*{6mm} Moments of angular distribution  $t_{M}^{L}$,  $p_{M}^{L}$ 
and $r_{M}^{L}$ are bilinear combinations of partial wave amplitudes. As 
described in the previous Section, we use
nucleon transversity amplitudes $g$ and $h$ corresponding to 
a given naturality exchange. It should be stresssed that the number of
 $t_{M}^{L}$ moments is smaller (see Table 1) than the number of real 
parameters characterizing the amplitudes, and therefore model-independent 
partial--wave analysis is not possible. 
Consequently, all  $\pi^{+}\pi^{-}$ phase 
shift studies that do not use polarized target data were based on non-trivial
physical assumptions like vanishing of spin-nonflip amplitudes (in our
language this corresponds to $g_{i}\equiv h_{i}$, where $i$ denotes $l,m$) in the unnnatural 
spin-parity amplitudes and phase coherence between the $m=0$ and $m=1$
amplitudes. It was shown in 
\cite{nowySvec,becker,becker_b} 
that these 
assumptions are badly broken by the polarized-target data. Unfortunately,
this fact has been  ignored in all subsequent \pipi  studies with
the notable exception of Svec papers
\cite{svec95,nowySvec}.  

On the other hand,  additional knowledge of  $p_{M}^{L}$ 
and $r_{M}^{L}$ moments yields the total number of observables which, as seen in Table 1,
slightly exceeds the number of real parameters. Since the 
present analysis is restricted to low $t$, we can ignore all $m>1$ 
amplitudes as all $M>2$ moments vanish (for a high--$t$ study see 
 \cite{RS}).\\ 
\hspace*{6mm}Terms appearing in amplitude combinations 
are of the type $|g_{i}|^2$,  $|h_{i}|^2$, $Re(g_{i}g_{j}^{*})$ or 
 $Re(h_{i}h_{j}^{*})$ but there is no mixed term like  $Re(g_{i}h_{j}^{*})$.
Consequently, we cannot determine the relative phase between the $g$ and
$h$ amplitudes in a model-independent way but the transversity amplitudes can
be completely determined {\it independently within each set}. 
This includes moduli and {\it relative phases} with the warning that 
all the relative phases within each set can be multiplied by $-1$.\\
\hspace*{6mm}In the analysis we expect to find some discreet ambiguities 
that arise from bilinearity of the equations.
In order not to lose any solution, great effort was devoted to providing  
many different starting points to the MINUIT program
\cite{JR}. 
They were as follows (see  \cite{LR} for details):\\
i) exact analytical solutions (possible for $l_{max}=1$ only),\\
ii) approximate analytical solution assuming that intensity of the $P_{0}$
wave is much smaller than that of the $S$ and $D_{0}$--waves (for $l_{max}=2$),
\\
iii) approximate analytical solution assuming phase coherence (for $l_{max}=2$
and $l_{max}=3$),\\
iv) several sets of random values,\\
v) results of the fit from neighbouring bins.\\
It should be stressed that approximations in ii) and iii) were
used for finding the starting values {\it only}; {\it it was always
the exact formulae that were fitted}. In  \cite{becker} such fits were 
done in 40 MeV bins for \\
\begin{center}
580 MeV $<$ \mpp $<$ 1780 MeV \\
0.01 GeV$^{2}$/c$^{2}$ $< |t| <$ 0.20 GeV$^{2}$/c$^{2}$.
\end{center}
Later,  similar analysis was performed in finer mass bins 
($\Delta$ \mpp = 20 MeV) in a slightly different kinematical region, i.e.\\   
\begin{center}
600 MeV $<$ \mpp $<$ 1600 MeV \\ 
0.005 GeV$^{2}$/c$^{2}$ $< |t| <$ 0.200 GeV$^{2}$/c$^{2}$.
\end{center}
\hspace*{6mm}The fits became more difficult and time-consuming as 
higher partial waves were included; therefore,  the MINOS error 
analysis (see  \cite{JR} for details) was required for all parameters 
for  $l_{max}=1$ only.
For higher masses, when more waves were needed, the MINOS errors were
calculated for the leading $m=0$ waves only. The solution was considered to
be acceptable only if the MINOS error analysis was possible; the $\chi^{2}$
values, being generally good, were  hardly helpful in selecting solutions.
Quite often there was
only one such solution in a given bin, although many different starting points
were used. 
However, in the mass region below $f_{0}(980)$ there are
two branches of solutions, best seen in the moduli of the S-wave g and h 
amplitudes around 900 MeV. This reflects the old "up-down" ambiguity. 

Intensities $I_{i}=|g_{i}|^{2}+|h_{i}|^{2}$ of partial
waves obtained in the 20 MeV analysis were published in  \cite{chabaud}, in fact
they still represent the most accurate measurements of the $f_{2}(1270)$
parameters
\cite{pdg96}. 
Moduli and relative phases of transversity amplitudes
are used in this paper for the first time.

 
\section{Results \label{numeric}}
\vspace{0.3cm}
 
{\bf a) Determination of the \mbox{\boldmath $S$}--wave}
 
\vspace{0.3cm}
\hspace{0.5cm}
In Figs 2--4 we show the experimental results obtained by the
CERN--Cracow--Munich collaboration as described in the previous Section.
Independent variables used in their partial wave analysis 
were: the sum $|\overline{g}|^2 + |\overline{h}|^2$ (Fig. 2a),
the ratio $|\overline{g}| / |\overline{h}|$ (Fig. 2b),
as well as the phase differences $\vartheta_{g}^L-\vartheta_{g}^{L^{'}}$ 
and $\vartheta_{h}^L-\vartheta_{h}^{L^{'}}$
($L, L'=S,P,D,F$) (Figs 3 and 4).
In our analysis we have assumed that phases of the partial waves 
 are described  mainly by phases of the  final state interaction amplitude 
of the \pipi system.
It means in particular that phases of the $P$, $D$ and $F$--waves follow  
phases of \ro, \fd and \rf decay amplitudes into \pipi. Our
PWA  has been done in three \mpp  effective mass regions
from 600 MeV to 1600 MeV.
 
\vspace{0.3cm}
 
{\em i) 600 MeV -- 980 MeV }
 
\vspace{0.3cm}
 
In this region one can safely  assume that only the $S$ and $P$--waves 
contribute since the $D$--wave is weak even at the upper limit of this region. 
It is only in this region that fully analytical solutions of the PWA
equations are possible \cite{LR}. 
The PWA analysis however, yields two 
solutions ("up" and "down") which are distinctly different in the \mpp 
region from 800 MeV to 980 MeV. In the "up" solution the sum 
$\mid \overline{g} \mid^2 + \mid \overline{h} \mid^2$ exhibits a maximum for 
\mpp $\approx$ 770 MeV, but in the "down" solution the moduli of 
$\mid \overline{g} \mid^2$ and  $\mid \overline{h} \mid^2$ are roughly 
constant from about 750 MeV to 980 MeV. The shape of the $S$--wave "up" 
modulus  was used by Svec to claim the existence of the \ep meson 
\cite{svec,svec95}.

$S$--wave phases of reaction (\ref{reaction_pol}) have been 
determined from phase differences between $S$ and $P$--waves.
In addition to the  "up--down" ambiguity in the moduli of the \g{} and \h{} 
transversity amplitudes, there is also a phase ambiguity in each \mpp bin.
This ambiguity comes from the mathematical structure of the PWA equations 
from which only cosines of the relative phases of the partial waves of 
reaction (\ref{reaction_pol}) can be obtained.
In our analysis we present two arbitrary choices of the $S$--wave phases
which form data sets shown in Figs 3 and 4.
In the first set, called "steep", $S$--wave phases grow faster than 
$P$--wave phases.
In the other set, called "flat", increase in $S$--wave phases is slower
than that for $P$--waves.Thus two sets of possible phases ("flat" or "down") 
combined with two branches 
of moduli ("up" and "down") lead us to consideration of four solutions for
the amplitudes g and h which can be called "up-steep","down-steep",
"up-flat" and "down-flat". Following this splitting of the amplitudes
also the $\pi-\pi$ phase shifts which will be derived from them will be
similarly labelled. In order to avoid a possible future confusion let us remark
here that in many papers in past the words "up" and "down" served mainly 
to distinguish the S-wave phase-shifts: the values of the phase shifts "up" 
were larger than the corresponding values of the solution "down". However, in 
the models in which the $a_1$ exchange has been
neglected and the S-wave amplitude was assumed elastic, the modulus of the 
"up" amplitude $\pi^{-}p \rightarrow \pi^{-}\pi^{+}n$
was {\it smaller} than the modulus of the amplitude "down" for the effective
$\pi^{-}\pi^{+}$ mass larger than about 800 MeV. The reason of this correlation
can be understood if we remember that the phase shifts of the "down"
solution were close to $90^{\circ}$ while the phase shifts of the "up"
solution were considerably larger in that effective mass range and the modulus
of the amplitude was proportional to $\mid \sin\delta \mid$. In our case, however,
we do not neglect an important contribution of the $a_1$ exchange, so we have
more combinations of the possible solutions for the amplitudes $g$ and $h$.

\vspace{0.3cm}
 
{\em ii) 980 MeV  -- 1460 MeV }
 
\vspace{0.3cm}
 
Here, the $S$, $P$ and $D$--waves have been included in the PWA analysis.
In this region the  \fd production is very strong  so the $D$--wave dominates 
and the $S$--wave phases have been evaluated from phase differences 
between the $S$ and $D$--waves.
In our model we have assumed that in the 980 MeV -- 1460 MeV region 
differences between the $P$ and $D$--wave are positive since the phase of the \ro
meson decay amplitude into two pions is larger than the corresponding phase
of the \fd meson amplitude. 
This allowed us to fix the sign of phase difference 
between the $S$ and $D$--waves in each energy bin, thus avoiding phase
ambiguity present in the first region.
 
\vspace{0.3cm}
 
{\em iii) 1460 MeV -- 1600 MeV }
 
\vspace{0.3cm}
 
Here, the $S$, $P$, $D$ and $F$  (dominated by the  \rf meson) waves have been 
included in the PWA analysis. Consequently,
the $S$--wave phases have been calculated from phase differences between 
the $S$ and $F$--waves. 
The sign of each  difference has  been fixed by the
assumption that phase  differences between the $P$ and $F$--waves as well as 
between the $D$ and $F$--waves are positive.
 
\vspace{0.3cm}
In each of the three regions production amplitudes of the \ro, \fd and 
\rf resonances follow the Breit-Wigner parametrisation of   
\cite{becker,chabaud},
namely
\be
M = \sqrt{A} \frac{m_{\pi\pi}}{\sqrt{q}}
 \frac{\sqrt{2l+1}m_R x_R\Gamma}{m_R^2 - m_{\pi\pi}^2-im_R\Gamma},\;\;\;\;\;\;\;
l = 1,2,3, 
\label{res_ampl}
\ee
where 
\be
\Gamma=\Gamma_R\left(\frac{q}{q_R}\right)^{2l+1}\frac{D_l(q_Rr)}{D_l(qr)},
\label{gamma_def}
\ee
and the Blatt--Weisskopf functions $D_l(qr)$ have the form:
\begin{eqnarray}
D_1(qr)&=&1+(qr)^2  \mbox{\hspace{4.8cm} for $P$--wave}, \nonumber\\
D_2(qr)&=&9+3(qr)^2+(qr)^4  \mbox{\hspace{3.2cm} for $D$--wave}, \nonumber\\
D_3(qr)&=&225+45(qr)^2+6(qr)^4+(qr)^6  \mbox{\hspace{1cm} for $F$--wave}.
\label{blatt_weisskopf}
\end{eqnarray}
In (\ref{res_ampl})  $A$ is a normalization constant , $m_R$, $\Gamma_R$ and
$x_R$ are mass, width and inelasticity of the resonance $R$ 
($R$=\ro, \fd or \rf),
$q$ is momentum of any pion in the \pipi rest system, and $q_R$ 
stands for momentum $q$ for \mpp = $m_R$.
Range parameter $r$ for \ro is equal to 4.8 GeV$^{-1}$ for the  "up" solution, 
and  5.3 GeV$^{-1}$ for the "down" solution 
\cite{chabaud}.
For the \fd and \rf resonances  $r$ equals 10 GeV$^{-1}$   and 3 GeV$^{-1}$ 
respectively
\cite{becker}.
 
Simultaneous presence of the $P$ and $D$--waves  for \mpp $>$ 980 MeV 
allowed us to check the validity of parametrisation of the 
$P$ and $D$--wave  phases by resonant amplitudes (\ref{res_ampl}).
In Fig. 5 we have shown phase differences between the $P$ and $D$ 
waves for \g{} and \h{} transversity amplitudes.
The $P$ and $D$--waves in the \h{} transversity amplitude are well described by 
the \ro and \fd resonant amplitudes  but in the case of the \g{} amplitude 
such parametrisation is not sufficient.

Since in the PWA analysis there is no phase relation between the
\g{} and \h{} transversity amplitudes, we have defined the $S$--wave
phases in the  following way:
\be
\vartheta_g^S=\left\{ \begin{array}{ll}
\vartheta_g^S-\vartheta_g^P + \theta_{\rho(770)}  
&
\mbox{for \hspace{.1cm} 600 MeV $\leq m_{\pi\pi} \leq$ 980 MeV}, \\
\vartheta_g^S-\vartheta_g^D + \theta_{f_2(1270)}+ \Delta
&
\mbox{for \hspace{.1cm} 980 MeV $\leq m_{\pi\pi} \leq$ 1460 MeV}, \\
\vartheta_g^S-\vartheta_g^F + \theta_{\rho_3(1690)}
&
\mbox{for \hspace{.1cm} 1460 MeV $\leq m_{\pi\pi} \leq$ 1600 MeV}, 
\end{array}
      \right.
\label{fazy_g_s}
\ee
 
\be
\vartheta_h^S=\left\{ \begin{array}{ll}
\vartheta_h^S-\vartheta_h^P + \theta_{\rho(770)} + \Delta
&
\mbox{for \hspace{.1cm} 600 MeV $\leq m_{\pi\pi} \leq$ 980 MeV}, \\
\vartheta_h^S-\vartheta_h^D + \theta_{f_2(1270)} + \Delta
&
\mbox{for \hspace{.1cm} 980 MeV $\leq m_{\pi\pi} \leq$ 1460 MeV}, \\
\vartheta_h^S-\vartheta_h^F + \theta_{\rho_3(1690)}
&
\mbox{for \hspace{.1cm} 1460 MeV $\leq m_{\pi\pi} \leq$ 1600 MeV} 
\end{array}
\right.
\label{fazy_h_s}
\ee
In  (\ref{fazy_g_s}) and (\ref{fazy_h_s}) $\theta_{\rho(770)}$, 
$\theta_{f_2(1270)}$ and $\theta_{\rho_3(1690)}$ are phases of the 
resonant amplitudes defined in (\ref{res_ampl}).
Function $\Delta$ has the form:
\begin{equation}
\Delta = \left\{ \begin{array}{cc}
50.37^{\circ} & \mbox{for 600 MeV $\leq$ \mpp $\leq$ 990 MeV},\\
50.37^{\circ}-0.116^{\circ}(m_{\pi\pi}-990\mbox{ MeV}) & \mbox{
for 990 MeV $\leq$ \mpp $\leq$ 1420 MeV},\\
0 & \mbox{\hspace{-2.15cm} for \mpp $\geq$ 1420 MeV}.
\end{array} \right.                                   \label{delta_g_function}
\end{equation}
Function $\Delta$ has been introduced in order to parametrize the
 differences between
$(\vartheta^P_h-\vartheta^D_h)$ and $(\vartheta^P_g-\vartheta^D_g)$.
We have determined this function empirically from  a linear fit 
in the range of 980 MeV $\leq$ \mpp $\leq$ 1300 MeV.
 
\vspace{0.3cm}
 
{\bf b) Separation of the $S$--wave into pseudoscalar- and pseudovector-exchange 
components }
 
\vspace{0.3cm}
Transversity amplitudes \g{} and \h{} have been averaged 
over $t$ as it was described in Sect. \ref{form}.
Coefficients $c_1, c_2, d_1$ and $d_2$ depend on averages 
$\overline{pU}$ and
$\overline{qV}$, and therefore they also depend on the form of functions $p(t)$ and
$q(t)$.
We have checked the dependence of amplitudes $\overline{g_{\alpha}}$, 
$\overline{h_{\alpha}}$ ($\alpha=A,B$)
and $A$, $B$ on the value of parameter $a$ (Eq. \ref{25}) for  0 $\leq$ $a$ $\leq$ 
4  GeV$^{-2}$. 
For both "up" and "down"  solutions and for both "flat" and "steep" sets of 
the $S$--wave phases below 1000 MeV, the differences in moduli and phases
of amplitudes
$\overline{g_{A}}, \overline{h_{A}}$ and $\overline{g_{B}}$, $\overline{h_{B}}$
are not higher than $3\%$ for 600 MeV $\leq$ \mpp $\leq$ 1600 MeV.
In the case of the $A$ amplitude the differences between different solutions 
can reach 10$\%$.
We have evaluated parameter $a$ to be 3.5 GeV$^{-2}$ 
from experimental data 
\cite{becker_b}
in the mass range of 710 MeV $<$ \mpp $<$ 830 MeV.
Since we do not know its mass dependence outside this region, we have kept it
fixed to 3.5 GeV$^{-2}$ in the whole \mpp range.
 
We have also checked the dependence of  pseudoscalar and pseudovector exchange 
amplitudes on the form of function $q(t)$ for the following functions: 
$q(t) = 1, t, e^{bt}$ and $q(t) = te^{bt}$, where $b=4~ $GeV$^{-2}$.
The second and fourth forms come from parametrization of the Regge propagator 
for the \aone exchange at $t \approx 0$
\cite{irving}.
Differences in moduli and phases of the $A$, $\overline{g_{A}},
\overline{h_{A}}$, $\overline{g_{B}}$ and $\overline{h_{B}}$ amplitudes 
caused by different functional forms of $q(t)$ are smaller than 5$\%$.
Very small changes of  amplitude $B$ (smaller than $0.5\%$)
are due to the strong factorization of the average
$\overline{qV} \simeq \overline{q}\overline{V}$.
Therefore, in the numerical calculations we have chosen $q(t) = 1$.
 
Parameter $f$ introduced in (\ref{38}) has an influence on amplitudes
$a_0$ and $a_S$ defined in Sect. \ref{form}. 
We have evaluated this parameter by
minimizing differences $1-\eta$ for 600 MeV $\leq$ \mpp $\leq$ 990 MeV for each 
solution ("up" and "down") combined with the phase set ("flat" and "steep") separately.
 
Since the dependence of transversity amplitudes \g{\alpha}
and \h{\alpha} ($\alpha=A, B$) on the shape of functions
$p(t)$ and $q(t)$ is weak, we  have used them to separate contributions of 
one--pion and \aone exchanges.
Amplitudes \g{\alpha} and \h{\alpha} depend on moduli and relative phases 
of amplitudes \g{} and \h{}.
Since for \mpp $<$ 980 MeV there are two solutions for moduli ("up" and "down")
and two solutions for relative phases ("flat" and "steep"), we obtain
four combinations of amplitudes which will be further labelled
\upp, \ups, \downp and \downs.
In Fig. 6 and 7 the moduli and phases of these four amplitudes are shown.
\vspace{0.3cm}

{\bf c) Properties of pseudoscalar- and pseudovector-exchange amplitudes  }
 
\vspace{0.3cm}
Separation of amplitudes leads to the equality of corresponding moduli:
$\mid \overline{h_{A}} \mid=\mid \overline{g_A} \mid$ and 
$\mid \overline{h_{B}} \mid=\mid \overline{g_B} \mid$
(see Eqs \ref{12} and \ref{13}).
However, phases of \h{A} and \h{B} differ from the corresponding phases of 
\g{A} and \g{B}.
Their differences, being functions of the effective mass, 
are related to the
phases of complex coefficients $c_i$ and $d_i$ ($i=1,2$ in
Eqs \ref{26}--\ref{33}).
The first phase difference, defined as the phase of $\overline{h_{A}}$  minus 
the phase of $\overline{g_A}$, increases monotonically
from about 10$^{\circ}$ to 50\crc in the effective mass range from 600 MeV to
1600 MeV.
The second phase difference, defined as the phase of $\overline{h_{B}}$  minus 
the phase of $\overline{g_B}$, is almost constant ($\approx -174^{\circ}$)
in the whole region of the effective mass.
This behaviour follows from the fact that 
the imaginary part of coefficient $U$ (Eq. \ref{14}) is much smaller 
than the real part of $U$, and the imaginary part of coefficient
$V$ dominates over its real part (Eq. \ref{15}).
Since we assume that functions $p(t)$ and $q(t)$ are real, the phase of
amplitude $A$ is the arithmetic average of the \g{A} and \h{A} phases
while that of amplitude $B$ is the arithmetic average of the 
\g{B} and \h{B} phases (compare Eqs \ref{12} and \ref{13}).
Their values differ strongly between four sets of solutions.
The \g{A} phases systematically  increase for the \pipi effective mass between 
600 MeV and about 980 MeV. 
Sudden phase change and strong reduction of the \g{A} moduli for \mpp
$\simeq$ 1000 MeV results from the appearance of the narrow resonance \fo.
Similarly decrease in the phases and moduli of amplitudes \g{A} for 
\mpp $>$ 1400 MeV can be related to the presence of another scalar resonance
of mass between 1400 MeV and 1500 MeV.
In most cases, errors of the phases and moduli of \g{A} are smaller than the 
corresponding errors for \g{B}.

 The \aone exchange 
amplitude \g{B}  amounts on average to about 20$\%$ of the pion exchange
amplitude  but around 1000 MeV and 1500 MeV it is 
of the same order as \g{A}.
A slight increase in the \g{B} modulus and a decrease in the \g{A} modulus
for the effective mass above 1420 MeV (seen in Figs 6 a,b and 7 a,b) can be 
related 
to the presence of a scalar resonance \fgg coupled to channels such as
$\pi\pi$, $\eta\eta$, $\eta\eta'$ and $4\pi$ (or $\sigma\sigma$)
\cite{amsler_close96}, \cite{bugg_hadron95} -- \cite{exp_groups1}.
On the basis of the $p\overline{p} \rightarrow 5\pi^0$ annihilation data,
S. Resag from the Crystal Barrel Collaboration 
\cite{resag_hadron95}
suggested a possible coupling of the \fgg to $\pi(1300)\pi$, where $\pi(1300)$ couples in turn 
to $(\pi\pi)_S\pi$ with $(\pi\pi)_S$ denoting
the $\pi\pi$ pair in the $I=0, S$--state.
The possible enhancement of the $\mid \overline{g_B} \mid$ amplitude
seen  in Figs 6 and 7 above 1400 MeV can be, however, attributed to another
decay  mode of \fgg, namely to a system $a_1\pi$, where 
$a_1 \rightarrow (\pi\pi)_S\pi$.
However, the small value of the partial decay width of \aone 
into $(\pi\pi)_S\pi$,
which -- according to the Particle Data Group 
\cite{pdg96} --
is equal to $0.3\%$ of the total \aone width, is questionable.
This number follows from just one analysis 
\cite{longacre}
related to the not well confirmed four--quark model of Jaffe
\cite{jaffe}.
The data gathered by the ACCMOR collaboration
\cite{accmor}
for reaction $\pi^-p \rightarrow 3\pi p$ indicate a much larger \aone 
coupling to $(\pi\pi)_S\pi$ (even several tens in percent of the total 
\aone width).
In reaction $\pi^-p \rightarrow \pi^+\pi^- n$ a possible subprocess is the
\aone exchange followed by interaction 
$\pi^- a_1 \rightarrow (\pi^+\pi^-)_S$ in the isoscalar state.
It seems  therefore the pseudovector exchange process can also contribute
to possible   production of the \fgg scalar resonance in the reaction
under consideration.
\vspace{0.3cm}
 
      {\bf d)  Pseudoscalar \mbox{\boldmath $I=0$}, \mbox{\boldmath $S$}--wave  amplitudes  }

\vspace{0.3cm}
Finally, let us  extract the $\pi\pi$ scalar--isoscalar amplitude 
$a_0$ from pseudoscalar amplitude \g{A}just separated from pseudovector
amplitude \g{B}. We use formulae (\ref{35}) and (\ref{37})--(\ref{42})
to obtain phase shifts \del and inelasticity coefficient \et corresponding 
to $a_0$.  In Fig. 8 we show the dependence of those parameters on the effective
mass for solutions labelled "steep" (corresponding to a fairly steep
behaviour of the phases of amplitudes \g{} and \h{}).
For both solutions (\ups and \downs) we observe a characteristic fast increase 
of phase shifts near 760 MeV (see Figs 8 a,b). Now all the solutions have to pass a test
connected with the fitting of the complex coefficient 
$f=\mid f \mid e^{i\varphi}$ to the requirement that the
inelasticity coefficient $\eta=1$ in the whole mass range from 
$m_{\pi\pi}=600$ MeV to the $K\overline{K}$ production threshold.\\
\hspace*{6mm}It appears immediately that the  the  "down-steep" solution
fails this test.  Such an attempt  leads 
to unnacceptably high $\chi^{2}/NDF=80/17$ value obtained in the best case
for $\mid f \mid=0.78$ and $\varphi=-8.4^{\circ}$. In addition, there is 
a strange behavior of $\eta$ i.e. a  clear violation of unitarity constraint 
below  $m_{\pi\pi}<720$ MeV and above  820 MeV while $\eta<1$ between these two 
values (see Fig. 8 c). It should be noted that if we changed normalization 
to keep $\eta\leq1$ in the whole mass range in question we would deal with 
a puzzling  large inelasticity between $m_{\pi\pi}=720$ MeV and 
820 MeV. Therefore, in our opinion   the "down-steep" solution is nonphysical 
and should  be excluded. \\
\hspace*{6mm}The situation is different  for the "up-steep" solution,  
for which we can find a constant factor  such that $\eta$ does not deviate
from unity ($\chi^{2}/NDF=15/17$) for $\mid f \mid=0.73$ and $\varphi=
-9.2^{\circ}$. However, also here we observe a puzzling 
behaviour with $\eta$ systematically exceeding unity for $m_{\pi\pi}<720$ MeV
and 
above $m_{\pi\pi}=820$ MeV but staying below 1 in the intermediate mass range.\\
\hspace*{6mm} On the other hand both "flat" solutions exhibit natural 
behavior with  $\eta$ fluctuating around unity. In addition, minimizing the
$\eta-1$ differences below the $K\overline{K}$ production threshold 
we  obtain reasonable values of $f$ close to 1, 
namely $\mid f \mid=0.89$, $\varphi=-4.4^{\circ}$ for the solution "up" and 
$\mid f \mid=0.84$, $\varphi=-17.8^{\circ}$ for the solution "down".
Thus we are left with two favoured, nicely behaving,  solutions
("up-flat" and "down-flat") and one, somewhat  queer, but still acceptable 
"up-steep" solution. Let us remark that  {\it only} the last solution is
consistent with the relatively narrow $f_{0}(750)$ meson
\footnote{Following the PDG convention we call this object $f_{0}(750)$ 
instead of $\sigma(750)$ used in the original papers.}
claimed by Svec \cite{svec95,nowySvec} on the basis of {\it moduli} of 
transversity amplitudes determined with the help of the 
polarized-target data.\\
\hspace*{6mm}In all solutions there is a sudden drop in $\eta$ for the
effective mass near 1000 MeV, caused by an opening of a new  $K\overline{K}$
channel. Another decrease of $\eta$ can be seen above 1500 MeV.\\
\hspace*{6mm}In Figs. 8 a,b and 9 a,b the phase shifts $\delta$ for all 
solutions have been compared with those obtained from the analysis of the
$\pi^{-}p\rightarrow \pi^{+}\pi^{-}n$ reaction on an {\it unpolarized}
target [20] (solution B), where separation of the $\pi$-exchange and
 $a_{1}$-exchange amplitudes was impossible. In Fig. 9 b in the mass
region from 600 MeV to about 1400 MeV we see only minor differences 
between phase shifts corresponding to the "down-flat" solution and
the results of [20]. In the "up-flat" solution around 900 MeV, however,
the values of $\delta$ are higher than those of ref.[20] by several
tens of degrees as seen in Fig. 9 a (a qualitatively similar  difference 
led many years ago to the "up-down" nomenclature). The
"up-steep" solution (see Fig. 8 a) with a narrow resonance is entirely 
different from the results of ref.[20]. Let us notice that the characteristic
"jump" of the "up-steep" phase shifts at 980 MeV is smaller 
(about $100^{\circ}$) than those in the  the "up-flat" ($\approx120^{\circ}$)
and "down-flat" ($\approx140^{\circ}$) solutions.\\ 
\hspace*{6mm} Finally let us comment on the behaviour of the phase shifts at 
higher effective masses, well above the $K\overline{K}$threshold. Here the 
different solutions are quite close each other. This follows from our 
assumption about the phase difference between the D- and S-waves (formulated 
in Sect. 4a ii) and rather small differences between the phases $\varphi$ of 
the factor $f$ (eq. 40).

A short comment on the comparison of errors obtained for the phase shifts in 
this study and errors obtained in \cite{grayer} is in order here. 
Errors in the partial wave analysis 
\cite{chabaud}
 have been obtained from the MINOS subroutine of the
MINUIT program used to fit data independently in each effective mass bin.
Although the errors of the B analysis in \cite{grayer} are smaller, they do not
include systematic effects corresponding to simplifying assumptions
like, for example, dominance of the s-channel nucleon helicity flip 
amplitude and phase coherence of unnatural spin-parity exchange 
amplitudes with helicity 0 and 1. In our analysis we avoid these
assumptions which
have been found to be badly broken by the polarized target data 
already in 
\cite{becker},\cite{becker_b},\cite{chabaud}.

Finally, let us note that the $I=2$, $S$--wave  amplitude $a_2$ plays an important 
role in the determination of amplitude $a_0$ (Eq. \ref{37}).
In particular, especially at high effective masses, it influences the inelasticity 
parameter \et very much. 
As described above, phases of amplitude 
$a_2$ have been obtained in the experiment on an unpolarized target 
\cite{hoogland}.
Thus, they contain some unknown errors related to the unknown contribution of 
the \aone exchange.
We feel, however, that those errors can be less important than other errors of
amplitude $a_S$ calculated from  \g{A} (Eqs \ref{38} and \ref{35}).

\vspace{0.3cm}
 
{\bf e)  Resonance interpretation of results }
 
\vspace{0.3cm} 

Let us now discuss behaviour of phase shifts and inelasticity coefficients
in terms of the scalar $I$=0 resonances, both 
already known and those newly postulated (see Figs 8 and 9).

Systematic increase in phase shifts below 1000 MeV can be related to the
existence of a broad \sig meson which we have found in \cite{klm}
and called $f_0(500)$. In \cite{klm} we analysed data obtained on
an unpolarized target.
Applying now the same model 
to the  \downp and \upp  solutions obtained for the polarized target we obtain 
parameters  of the $f_0(500)$ meson very similar to those given in \cite{klm}. 
In this way, presence of 
$f_0(500)$ is reinforced. We should mention here that
N. A. T\"ornqvist and M. Roos \cite{torn96}
also support the existence of the \sig meson naming it
tentatively $f_0(400-900)$. The \sig meson of mass
555 MeV has also been reported recently by Ishida et al. (\cite{ishida}) in an 
analysis of the $\pi \pi$ phase shifts especially near the pion-pion
threshold. The threshold behaviour of the scalar $I$=0 scattering amplitudes
is very much influenced by the presence of the $f_0(500)$ state (see \cite{kal}).
The \sig meson plays also a very important role in nuclear physics
and in the description of nucleon-nucleon interactions. For example, in
\cite{bonn} typical value of 550 MeV has been used for its mass, which is in agreement
with the above findings. 

\hspace*{6mm}Since our "up-steep" solution (see Sec. 5d) corresponds to the 
narrow $f_{0}(750)$ claimed by Svec on the basis of (mainly) the same
data let us comment on the similarities and differences of two approaches.
Svec uses an "analytical solution+Monte Carlo" method to extract amplitudes 
from experimental moments while our amplitudes come from the standard
$\chi^{2}$ minimization method. The advantage of the latter is an
applicability to higher mass range where more partial waves contribute and
analytical solution is not possible. In the very recent draft\cite{nowySvec}
 he compares both
methods for $m_{\pi\pi}<900$ MeV finding an excellent agreement between two
methods. But there is an important  difference in the
further analysis. Svec finds narrow $f_{0}(750)$ in {\it only one
of two} transversity amplitudes and argues against summing the squares of 
two amplitude moduli, which - according to him - distorts or even hides the 
$f_{0}(750)$. 
In our analysis we extract the dominating $\pi$-exchange amplitude and 
weaker but non-negligible  $a_{1}$-exchange amplitude from connecting
{\it both} transversity amplitudes. Studying the  {\it pure} $\pi$-exchange 
amplitude we  find  one, less favoured but still possible, solution
exhibiting the narrow $f_{0}(750)$. A simple Breit-Wigner fit with the
energy-independent background yields $m=(770\pm8)$ MeV and 
$\Gamma=(120\pm9)$ MeV, well consistent with  $m=(753\pm19)$ MeV and 
$\Gamma=(108\pm53)$ MeV in Ref. \cite{nowySvec}. This solution, however,
cannot be well described by 
the above-mentioned model of Ref. [2] without a substantial modification of the 
pion-pion interaction.

As can be seen in Fig. 9 c, a sudden drop in inelasticity for the 
effective mass near 1000 MeV is caused by the opening of a new \KK channel.
This fact, along with a characteristic jump of phase shift \del coupled
with a rapid decrease 
in elasticity \et near 1000 MeV (Fig. 9 a,b), is due to the narrow
\fo resonance \cite{klm,pdg96}. 
We stress that this jump, although smaller than in two "flat" solutions,
is seen {\it also in the ``up-steep'' solution}, 
since in some older studies the ``up'' solution was rejected as inconsistent
with  the narrow \mbox{\fo.} Analysing the polarized target data Svec has shown that
an intensity of one S-wave transversity amplitude can be simultaneously
parametrized in terms of two interfering resonances $f_{0}(750)$ and \fo
 \cite{nowySvec}.
 
Another decrease in \et near 1500 MeV -- 1600 MeV can be related to the opening
of further channels like $4\pi$ ( $\sigma\sigma$ or $\rho\rho$ ),
$\eta\eta'$ or $\omega\omega$.
For \mpp larger than 1470 MeV, phase shifts for both solutions (\upp
and \downp) show a systematically steeper increase than phase  shifts  
corresponding to data obtained on the unpolarized target.
Both facts may be related to the possible existence of the \fgg resonance
\cite{amsler_glasgow,amsler_close96}, \cite{bugg_hadron95}--\cite{exp_groups1}.

 
\section{Summary \label{summary}}
 
\hspace{0.6cm}
A formalism permitting  extraction and separation of the $S$--wave 
pseudoscalar and pseudovector exchange amplitudes in
the reaction $\pi^- p_{\uparrow} \rightarrow \pi^+ \pi^- n$ on a transversely 
polarized target has been presented. A new analysis of the CERN--Cracow--Munich
 collaboration data obtained with 20 MeV bins at 17.2 GeV/c $\pi^-$ beam 
momentum on the polarized target has been performed in the \pipi energy range 
from 600 MeV to 1600 MeV. Already at the level of moduli this analysis yields
two solutions between 600 MeV and 980 MeV (the ``up-down'' ambiguity).

Due to the lack of information on the relative phases between transversity 
amplitudes \g{} and \h{} we have assumed that the relative phases of both
$S$--wave transversity amplitudes $g$ and $h$ are generally governed by the 
phase behaviour of the dominant resonant $P$, $D$ and $F$ partial wave 
amplitudes corresponding to the $\rho(770)$, $f_2(1270)$ and $\rho_3(1690)$ 
resonances decaying to the \pipi pairs and interfering with the $S$--wave.
This leads to an additional twofold ambiguity since the relative phases 
can be either added or subtracted. Thus we have "down-flat", "down-steep",
"up-flat" and "up-steep" solutions. However the "down-steep" solution is shown 
to violate unitarity and can
be ignored. The remaining three solutions are acceptable although the 
the "up-steep" one exhibits a peculiar behavior of inelasticity.

The \aone exchange amplitudes are especially important at 1000 MeV and 1500 MeV 
and cannot be neglected with 
respect to the $\pi$ exchange amplitudes.
 This puts in serious doubt all the PWA 
results which assumed absence of the \aone exchange.\\
\hspace*{6mm}
Separation of the $\pi$--exchange from the $a_1$--exchange allowed us to 
calculate the $I=0$, $S$--wave $\pi\pi$ amplitudes in a weakly--model--dependent manner.
In the low--mass region both "flat" solutions are consistent with $f_{0}(500)$ while the 
phase behavior of the "up-steep" one agrees with the narrow $f_{0}(750)$.
Up to the energy of about 1420 MeV, phase shifts of our "down-flat"
solution agree within the errors with the one obtained without the 
polarized-target data. However, above 1420 MeV  the phase shifts 
in all solutions increase with
energy faster than those obtained without the polarized-target data. This
phase behavior as well as an increase  of the $a_{1}$-exchange amplitude 
can be due to the presence of the  $f_{0}(1500)$.

\vspace{0.5cm}
 
{\em Acknowledgements.} We thank L. G\"orlich, J. Kwieci\'nski, B. Loiseau, 
M. R\'o\.za\'nska and J. Turnau for enlightening discussions and communications.
We are also grateful to Dr. Svec for sending us his papers prior to
publication and for calling our attention to the  narrow $f_{0}(750)$
meson.

This work has been partially supported by the Polish State Committee for Scientific
Research (grants No 2 P03B 231 08, 2 P03B 043 09) and by the Maria 
Sk\l{}odowska--Curie Fund II (No PAA/NSF--94--158).  
 
 

\newpage

\begin{center}
FIGURE CAPTIONS\\
\end{center}

\addvspace{1cm}

\begin{itemize}

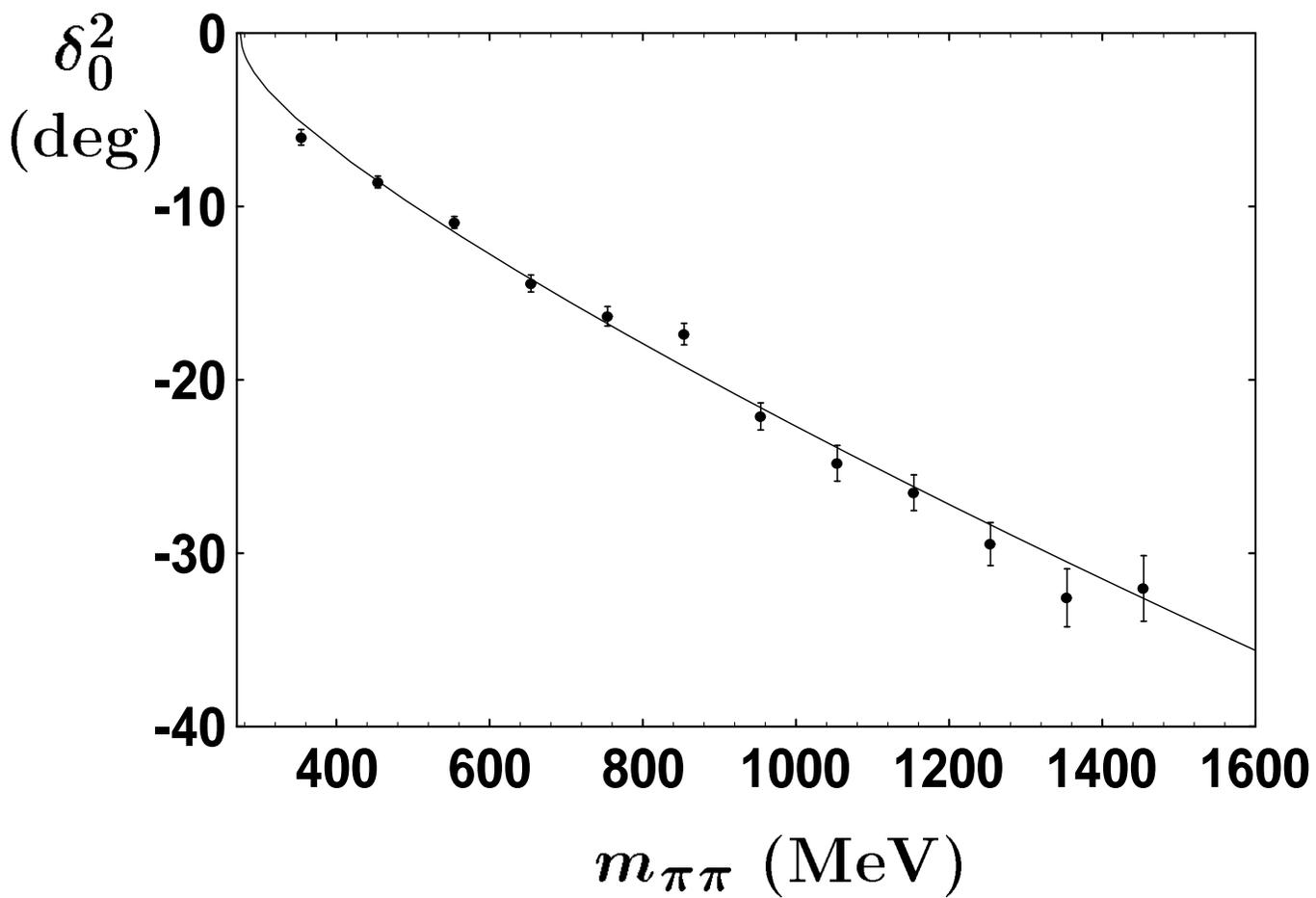
\item Fig.~1. The $I=2$, $S$--wave $\pi\pi$ phase shifts versus the
 effective \pipi mass.
The curve represents the fit to the data [32].
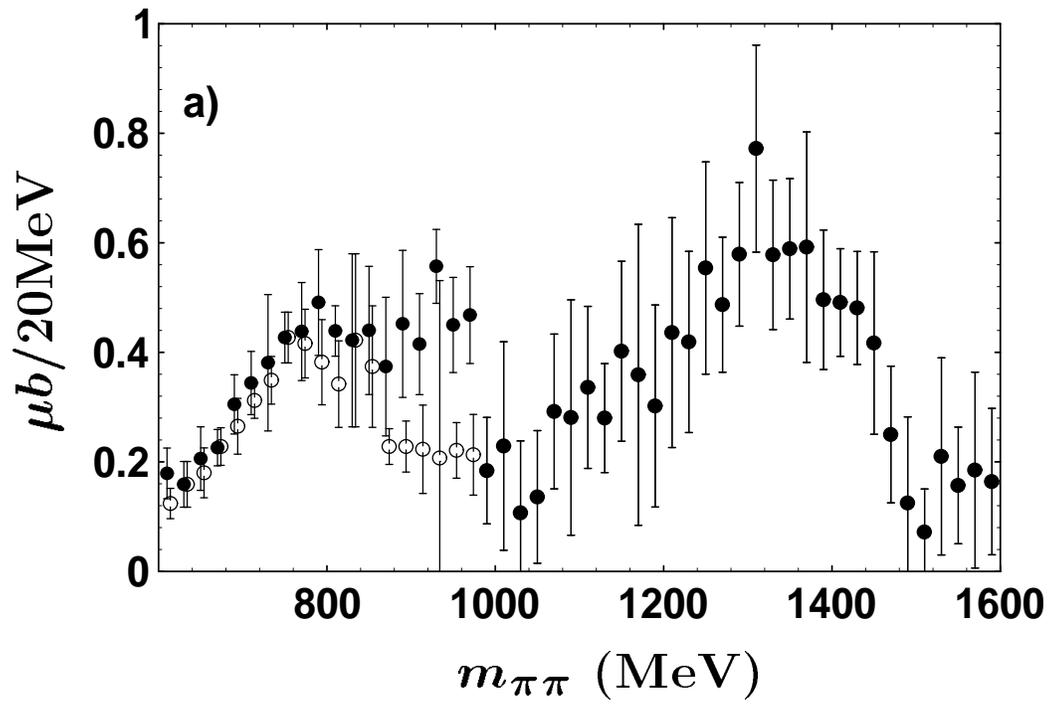
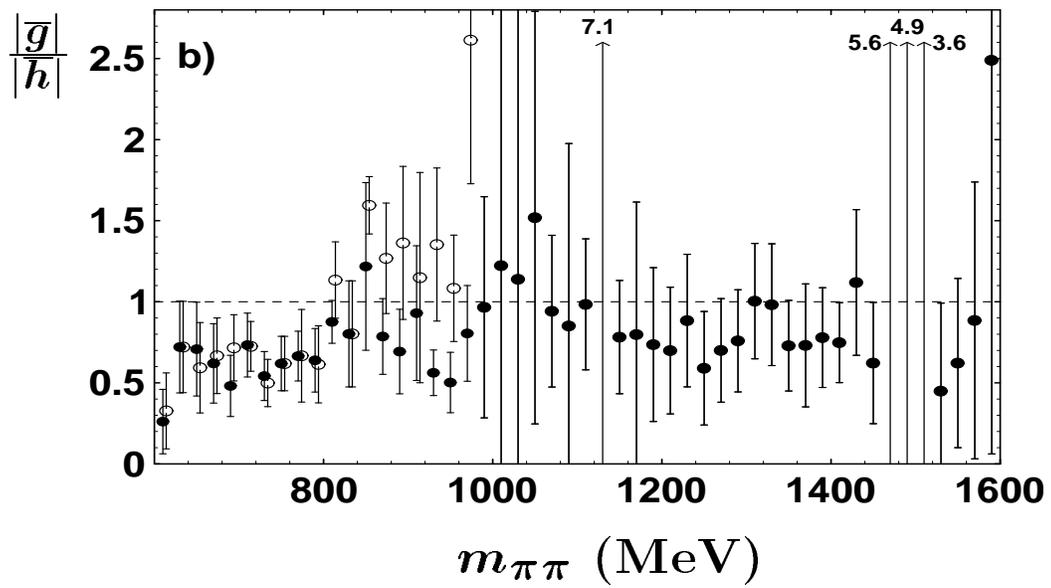
\item Fig.~2. Results of the partial wave analysis of the CERN--Cracow--Munich 
collaboration:
{\bf  a)} $\mid \overline g \mid^2 + \mid \overline h \mid^2$,
{\bf  b)} ratio $\mid \overline g \mid/\mid \overline h \mid$.
Below 980 MeV, full and open circles represent the "down" and "up" solution,
 respectively.
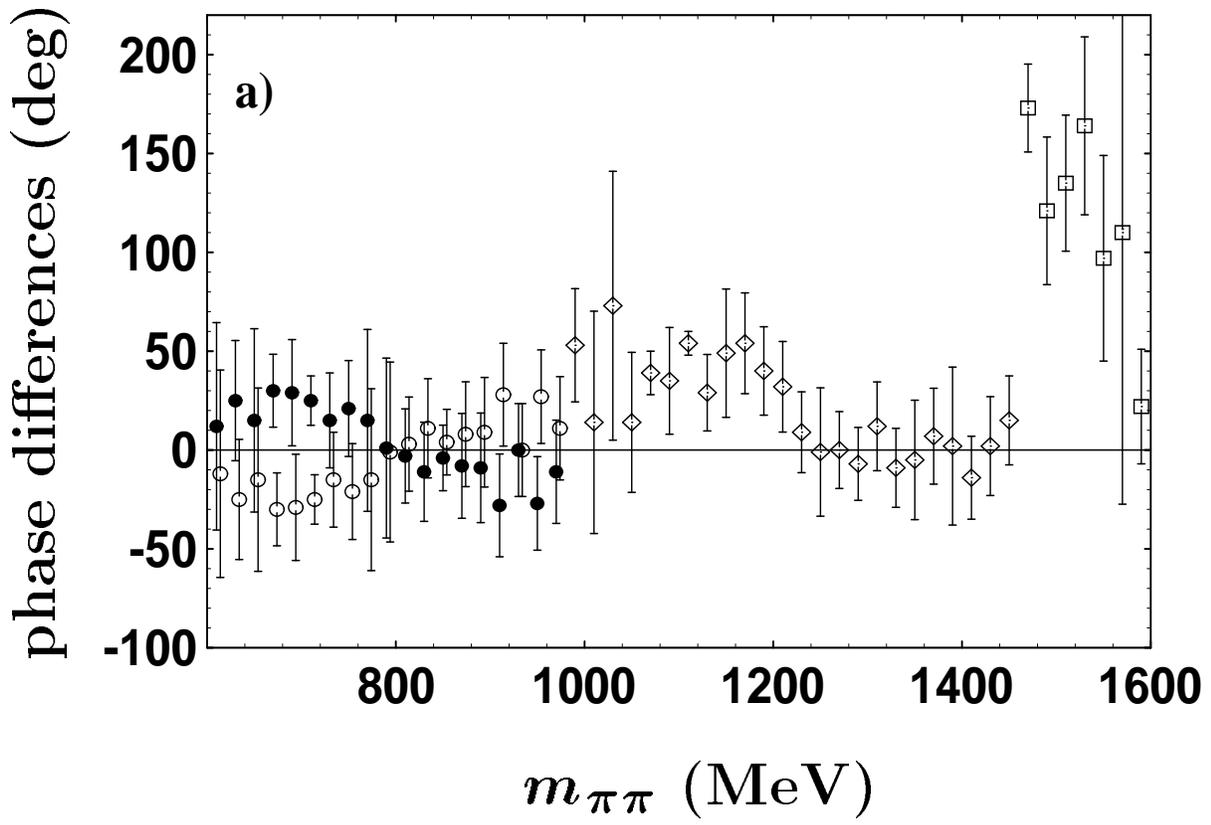
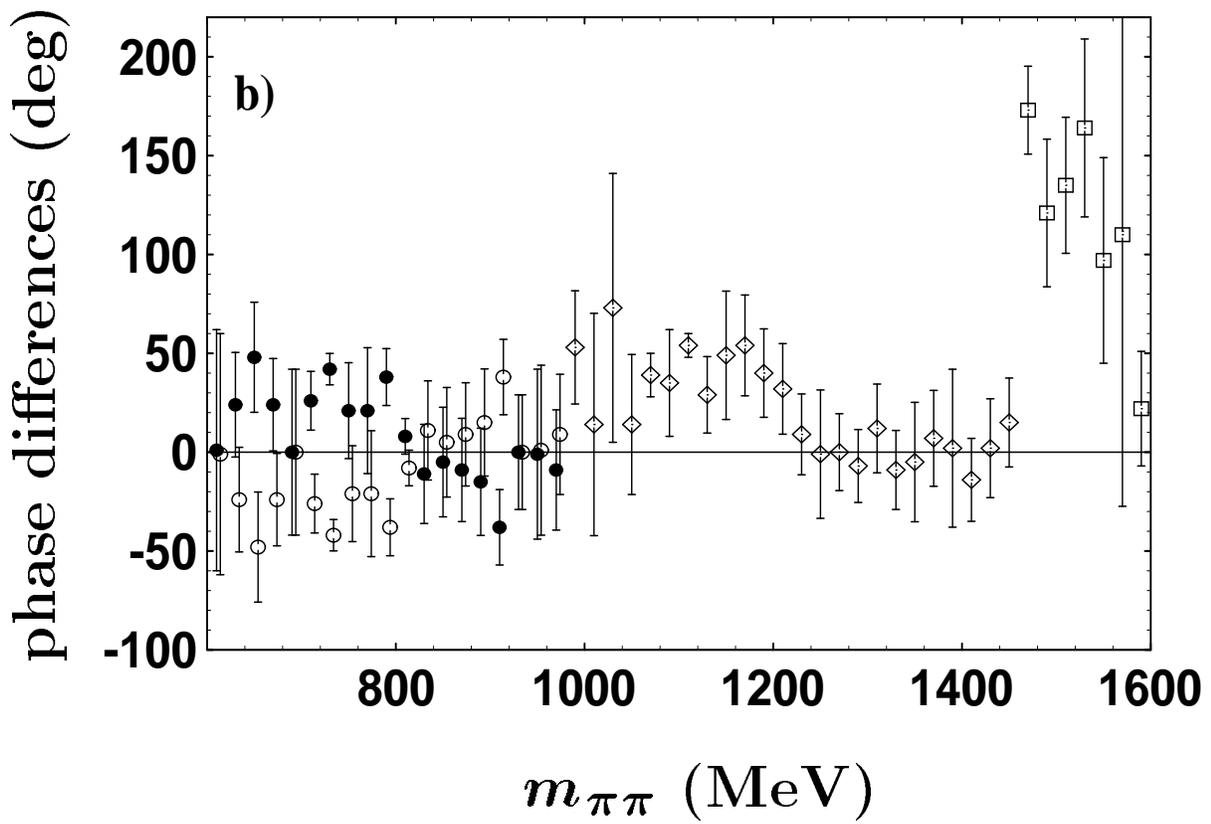
\item Fig.~3. Phase differences of the $g$ transversity amplitudes obtained
 in the 
partial wave analysis of the CERN--Cracow--Munich collaboration:
{\bf  a)} Solution "up": phase differences $\vartheta_g^S-\vartheta_g^P$ 
between the $S$ and $P$--waves  for the "flat" set  (full circles) 
and the "steep" set (open circles), 
phase differences $\vartheta_g^S-\vartheta_g^D$ between the $S$ and $D$--waves (diamonds) and 
phase differences $\vartheta_g^S-\vartheta_g^F$ between the $S$ and $F$--waves (squares).
{\bf  b)} Solution "down": notation as in a).
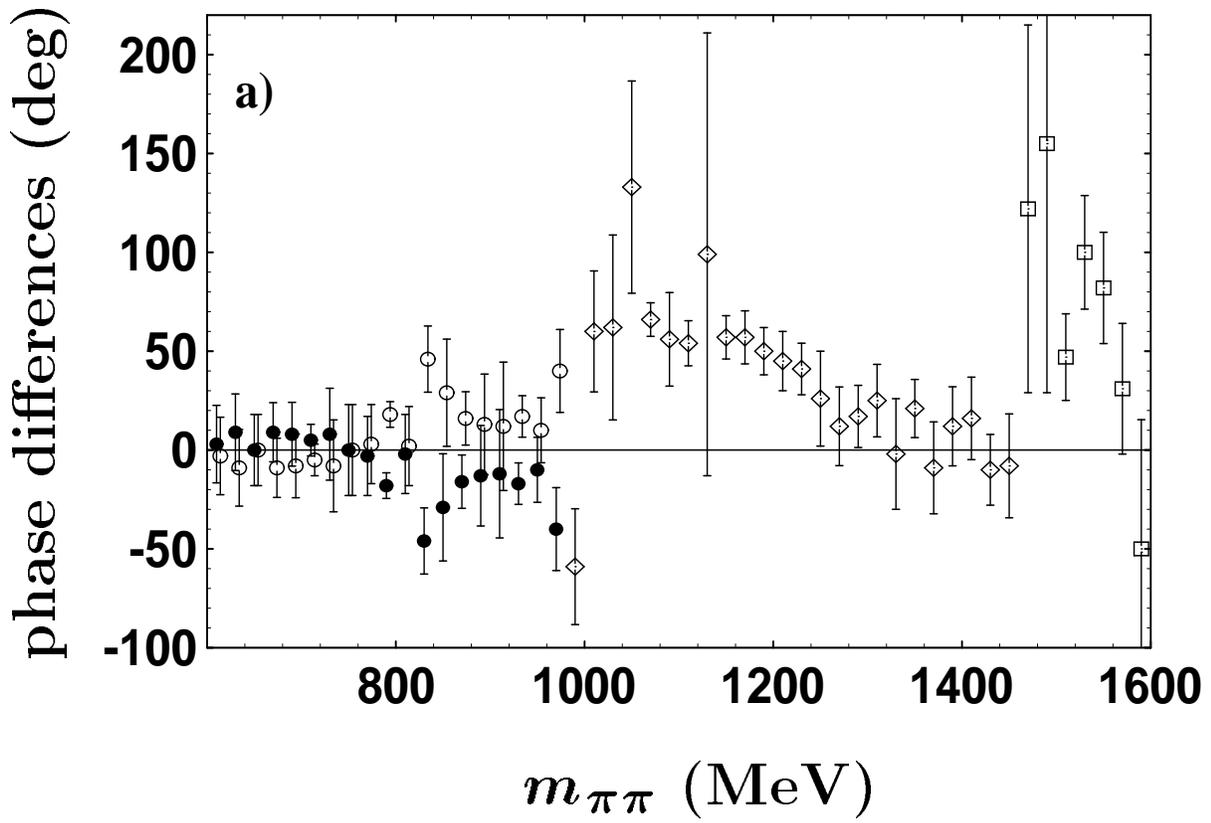
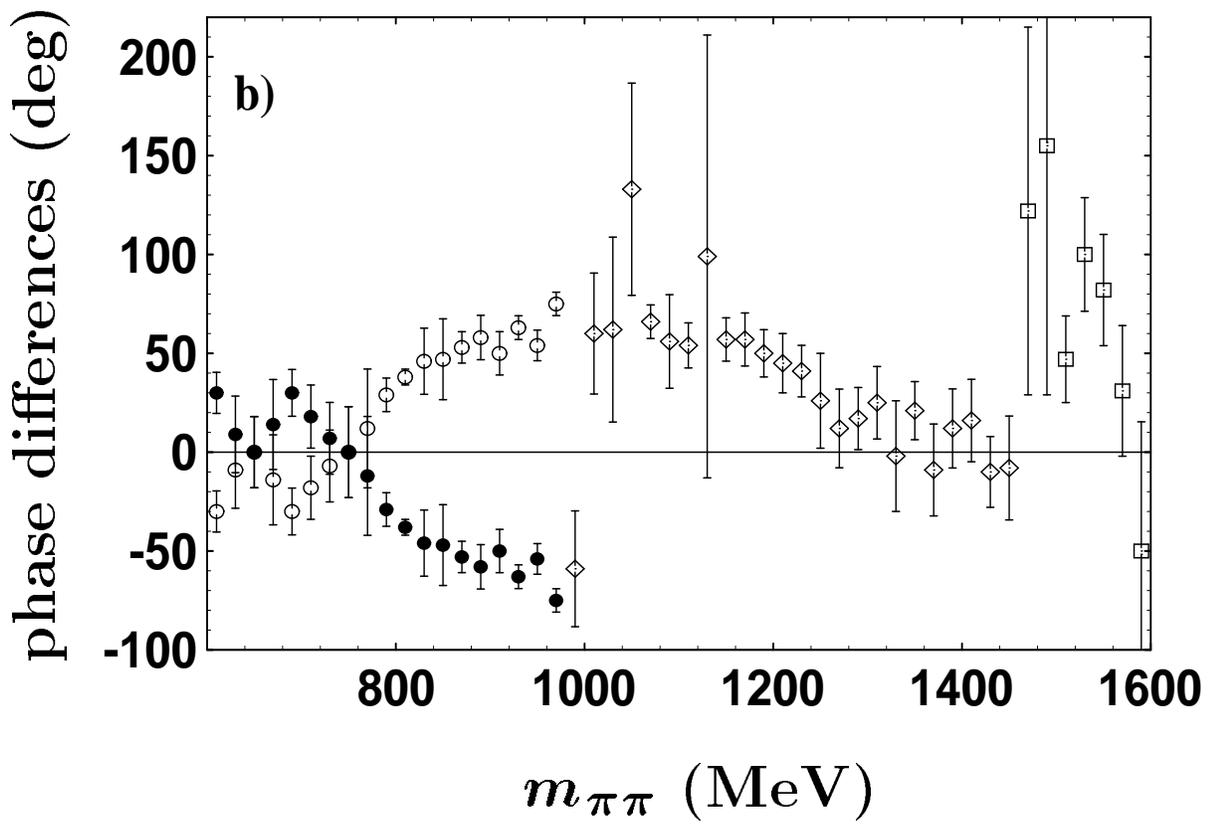
\item Fig.~4. Phase differences of the $h$ transversity amplitudes obtained in the 
partial wave analysis of the CERN--Cracow--Munich collaboration.
Notation as in  Fig. 3.
{\bf a)} Solution "up".
{\bf b)} Solution "down".
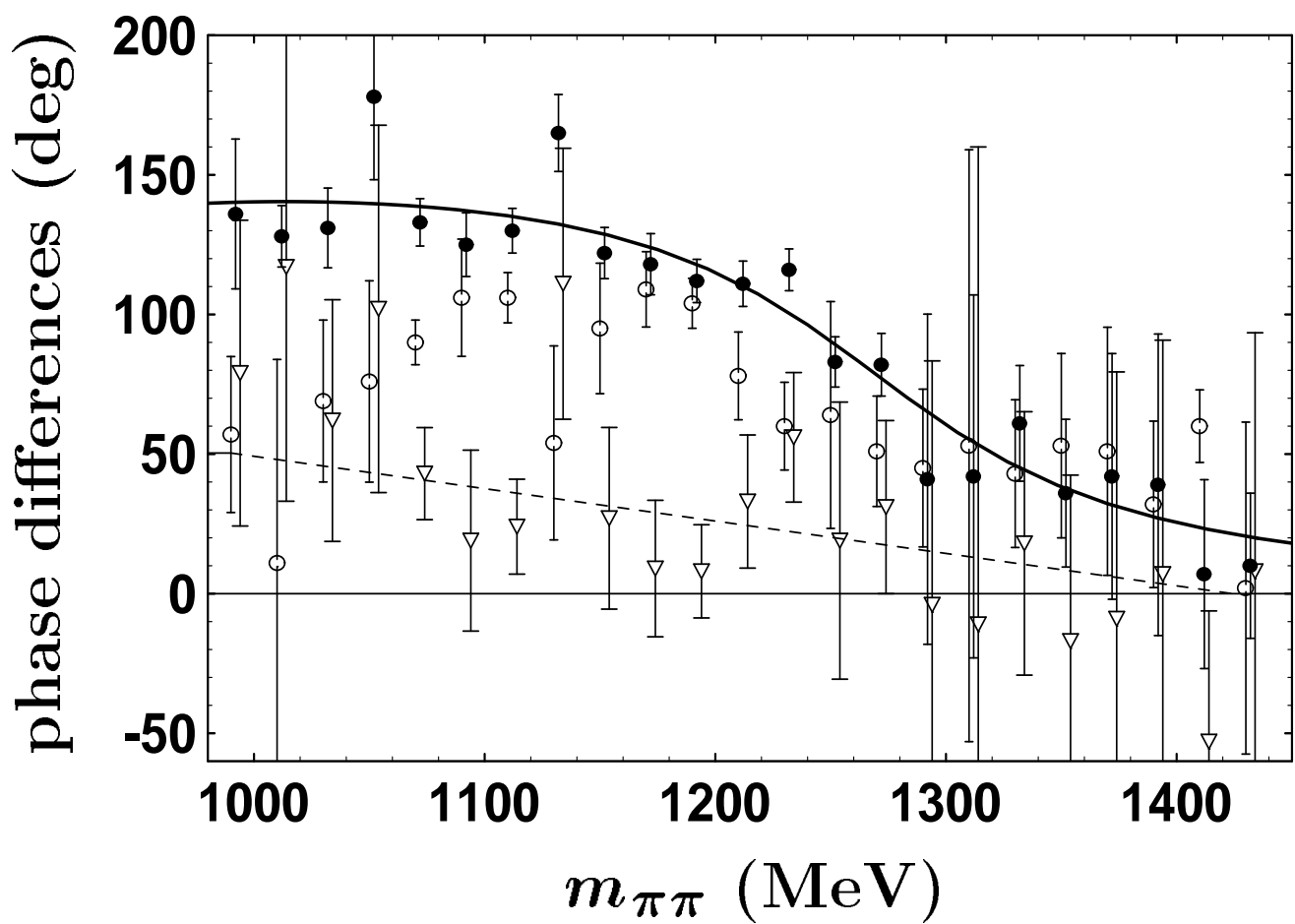
\item Fig.~5. Phase differences between the $P$ and $D$--waves for the 
$g$ ($\vartheta_g^P-\vartheta_g^D$: open circles) and  
$h$ ($\vartheta_h^P-\vartheta_h^D$: full circles) 
transversity amplitudes versus the effective \pipi mass.
Dashed line represents the effective \pipi mass dependence of function $\Delta$ obtained
from a fit to 
differences $(\vartheta_h^P-\vartheta_h^D) - (\vartheta_g^P-\vartheta_g^D)$
denoted by triangles.
Solid line represents differences between phases of the \ro  and 
\fd decay amplitudes.
\item Fig.~6. a) Moduli of pseudoscalar $\mid \overline{g_A} \mid
=\mid \overline{h_A} \mid$ (open circles)
and pseudovector $\mid \overline{g_B} \mid=\mid \overline{h_B} \mid$ 
(diamonds) exchange amplitudes as a function of the effective \pipi mass for
 the \upp solution.
{\bf b)} Same as in a) for the \ups solution.
{\bf c)} Phases of pseudoscalar exchange amplitudes \g{A} (open 
circles) 
and pseudovector amplitudes $\overline{g_B}$ 
(diamonds)  versus the effective \pipi mass for the \upp solution.
{\bf d)} Same as in c) for the \ups solution. 
{\bf e)} The phases of the pseudoscalar exchange amplitudes \h{A} (open 
circles) 
and the pseudovector amplitudes $\overline{h_B}$ 
(diamonds)  versus the effective \pipi mass for the \upp solution.  
{\bf f)} Same as in e) for the \ups solution.
\item Fig.~7. a) Moduli of pseudoscalar 
$\mid \overline{g_A} \mid=\mid \overline{h_A} \mid $ (full circles)
and pseudovector $\mid \overline{g_B} \mid=\mid \overline{h_B} \mid $ 
(diamonds) exchange amplitudes as a function of the effective \pipi 
mass for the \downp solution. 
{\bf b)} Same as in  a) for the \downs solution.
{\bf c)} Phases of pseudoscalar exchange amplitudes $\overline{g_A}$ 
(full circles) and pseudovector amplitudes $\overline{g_B}$ 
(diamonds) versus the effective \pipi mass for the \downp solution.
{\bf d)} Same as in  c) for the \downs solution.
{\bf e)} Phases of  pseudoscalar exchange amplitudes $\overline{h_A}$ 
(full circles) and  pseudovector amplitudes $\overline{h_B}$ 
(diamonds) versus the effective \pipi mass for the \downp solution.
{\bf f)} Same as in  e) for the \downs solution.
\item Fig.~8. {\bf a)} Scalar--isoscalar \pipi phase shifts 
$\delta_0^0$ as a function of the
 effective \pipi mass
for the \ups  solution
(open circles) and for data 
[20]
(triangles).
{\bf b)} Same as in a) for the \downs  solution (full circles).
{\bf c)} Scalar--isoscalar \pipi inelasticity coefficient
\et versus the effective \pipi mass for the \downs (full circles) 
and \ups (open circles) solutions.
\item Fig.~9. {\bf a)} Scalar--isoscalar \pipi phase shifts 
$\delta_0^0$ as a function 
of the effective \pipi mass
for the \upp  solution
(open circles) and for data 
[20]
(triangles).
{\bf b)} Same as in  a) for the  \downp solution (full circles).
{\bf c)} Scalar--isoscalar \pipi inelasticity coefficient
\et versus the effective \pipi mass for the \downp (full circles) 
and \upp (open circles) solutions.

\end{itemize}

 
%
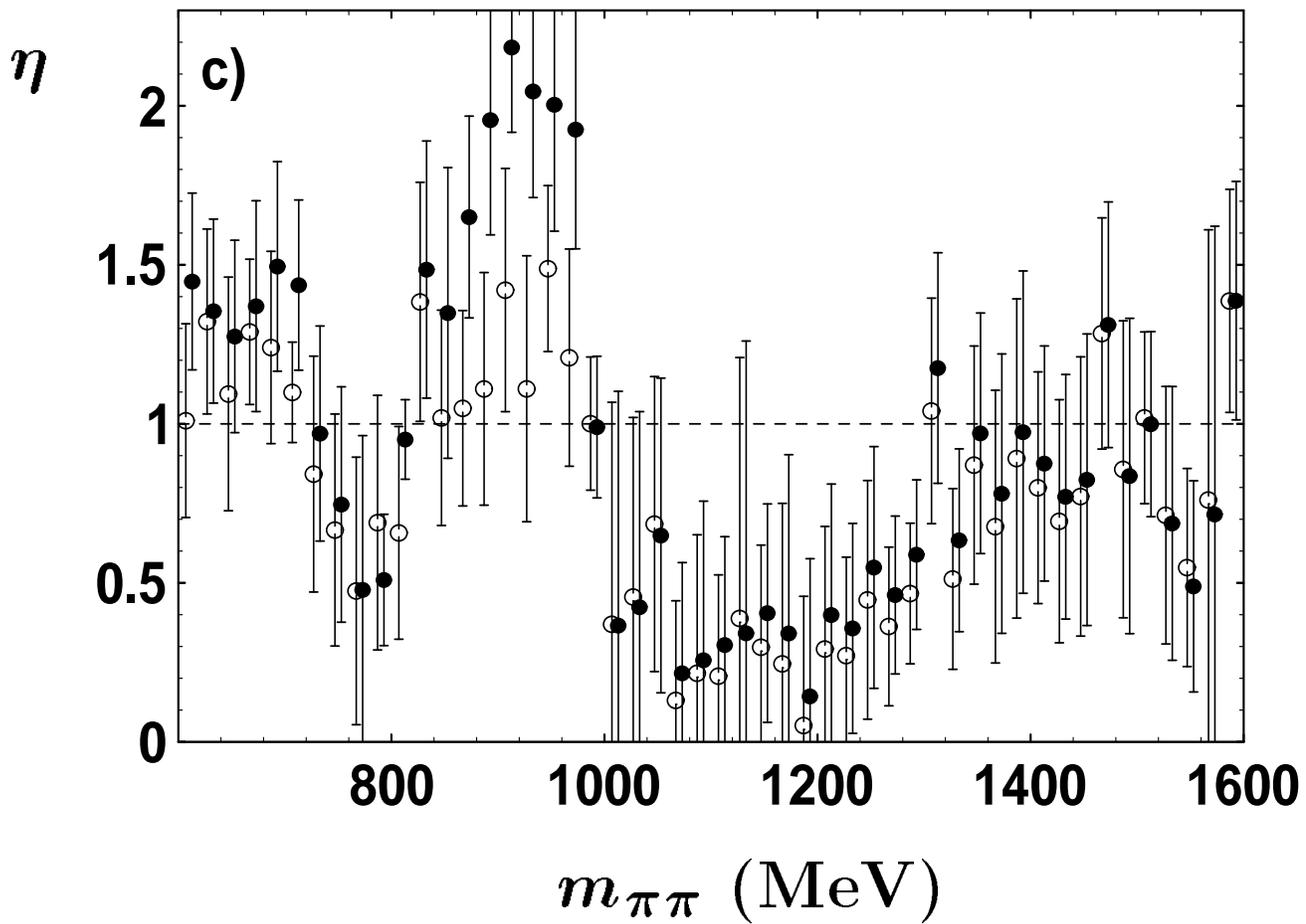
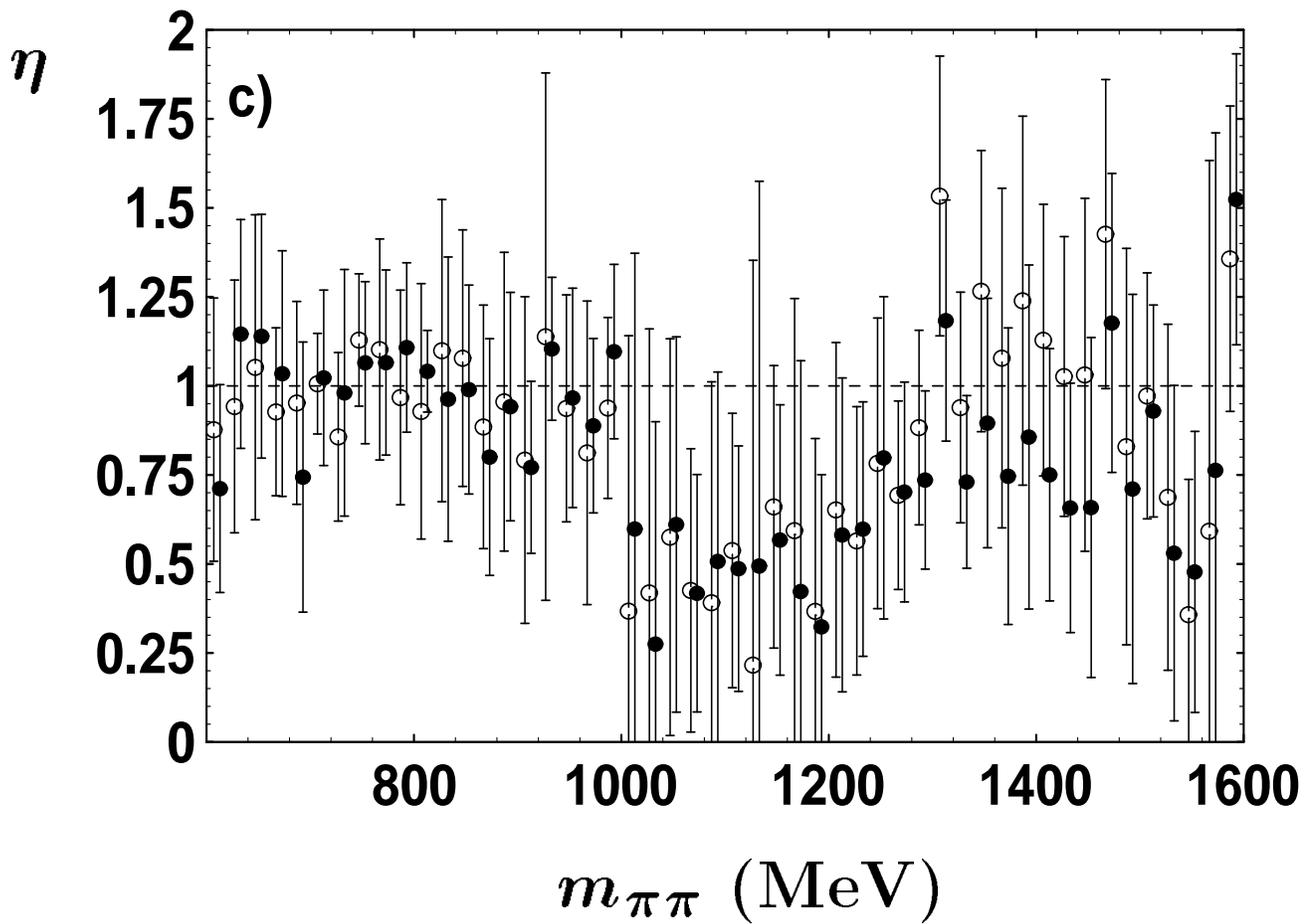
\end{document}